\documentclass[11pt]{article}
\usepackage{a4,amsfonts,amsmath,chicago,epsfig,fullpage}
\usepackage{calc,graphicx}

\newcommand{\bld}[1]{\mathbf{#1}}

\newcommand{\bb}[1]{\mathbb{#1}}

\newcommand{\mcal}[1]{\mathcal{#1}}
\newcommand{\bs}[1]{\boldsymbol{#1}}

\def\norm#1{\left\|#1\right\|}

\begin{document}

\begin{titlepage}

\title{A Multiresolution Census Algorithm for Calculating Vortex
  Statistics in Turbulent Flows}

\author{
Brandon Whitcher\thanks{Clinical Imaging Centre, GlaxoSmithKline,
  Hammersmith Hospital, Imperial College London, Du Cane Road, London
  W12~0HS, United Kingdom. E-mail:
  \textit{brandon.j.whitcher@gsk.com}.}
\and
Thomas C. M. Lee\thanks{Corresponding author.  Department of
  Statistics, The Chinese University of Hong Kong, Shatin, Hong Kong,
  and Department of Statistics, Colorado State University, Fort
  Collins, CO 80523-1877, United States.  E-mail:
  \textit{tlee@sta.cuhk.edu.hk}.}
\and
Jeffrey B. Weiss\thanks{Department of Atmospheric and Oceanic
  Sciences, University of Colorado, Boulder, CO 80309.  E-mail:
  \textit{jeffrey.weiss@colorado.edu}.}
\and
Timothy J. Hoar\thanks{Geophysical Statistics
  Project, National Center for Atmospheric Research, Boulder, CO
  80307-3000.  E-mail: \textit{thoar@ucar.edu},
  \textit{nychka@ucar.edu}.}
\and 
Douglas W. Nychka$^\S$
}

\date{April 24, 2006; revised: March 26, 2007}

\maketitle

\begin{abstract}
% The fundamental equations that model turbulent flow are difficult to
% evaluate directly.  
The fundamental equations that model turbulent flow do not provide
much insight into the size and shape of observed turbulent structures.
We investigate the efficient and accurate representation of structures
in two-dimensional turbulence by applying statistical models directly
to the simulated vorticity field.  Rather than extract the coherent
portion of the image from the background variation, as in the
classical signal-plus-noise model, we present a model for individual
vortices using the non-decimated discrete wavelet transform.  A
template image, supplied by the user, provides the features to be
extracted from the vorticity field.  By transforming the vortex
template into the wavelet domain, specific characteristics present in
the template, such as size and symmetry, are broken down into
components associated with spatial frequencies.  Multivariate multiple
linear regression is used to fit the vortex template to the vorticity
field in the wavelet domain.  Since all levels of the template
decomposition may be used to model each level in the field
decomposition, the resulting model need not be identical to the
template.  Application to a vortex census algorithm that records
quantities of interest (such as size, peak amplitude, circulation,
etc.) as the vorticity field evolves is given.  The multiresolution
census algorithm extracts coherent structures of all shapes and sizes
in simulated vorticity fields and is able to reproduce known physical
scaling laws when processing a set of voriticity fields that evolve
over time.
\end{abstract}

\bigskip


\end{titlepage}

\newlength{\ddpara}
\setlength{\ddpara}{1.5\baselineskip}
\baselineskip\ddpara

\section{Introduction}

% The large-scale fluid motion of planetary atmospheres and oceans is
% extremely turbulent and is strongly influenced by the planetary
% rotation and the planet's gravitational field.  The planetary
% rotation and gravitation render the resulting fluid motion primarily
% horizontal.  For example, in the Earth's atmosphere, vertical motion
% is typically at speeds of several~cm/s, while strong horizontal
% motions such as the jet stream can exceed speeds of 100~m/s.  As a
% result of these influences, planetary atmospheres and oceans are
% self-organized into coherent structures
% {\cite{mcc-wei:anisotropic}}.  The two main categories of coherent 
% structures are vortices; such as hurricanes, tornados, oceanic
% vortices, and Jupiter's Great Red Spot, and jets; such as the
% Earth's atmospheric Jet Stream, and the Gulf Stream in the North
% Atlantic Ocean.  With respect to the scope of this manuscript we are
% only interested in identifying vortices, so the term coherent
% structure will be synonymous with vortex.

The large-scale fluid motion of planetary atmospheres and oceans is
extremely turbulent and is strongly influenced by the planetary
rotation and the planet's gravitational field.  The planetary rotation
and gravitation render the resulting fluid motion primarily
horizontal.  For example, in the Earth's atmosphere, vertical motion
is typically at speeds of several~cm/s, while strong horizontal
motions such as the jet stream can exceed speeds of 100~m/s.
Turbulent fluids are characterized by having a wide range of spatial
scales with complex non-linear interactions between the scales.  One
noteworthy feature of turbulent fluids is that they self-organize into
coherent features.  The two main categories of large-scale coherent
structures are vortices; such as hurricanes, tornados, oceanic
vortices, and Jupiter's Great Red Spot, and jets; such as the Earth's
atmospheric Jet Stream, and the Gulf Stream in the North Atlantic
Ocean.  With respect to the scope of this manuscript we are only
interested in identifying vortices, so the term coherent structure
will be synonymous with vortex.

The reason for the formation of coherent structures is poorly
understood. Due to the quasi-horizontal nature of atmospheres and
oceans, the energy cascades from small to large scales, and the
accumulation of energy at large scales is associated with large-scale
coherent structures {\cite{mcc-wei:anisotropic}}.  Structures may also
be formed from the growth of instabilities in the flow, with the scale
of the structure determined by the scale of the instability.
Regardless of their formation mechanism, accepting that such
structures exist in turbulent flows and analyzing their behavior and
impact has led to significant advances in understanding turbulence.

In many instances we wish to know the statistics of vortex
properties. While traditional theories of turbulence are framed in
terms of energy spectra, more recent theories are based around the
statistics of the vortex population
{\shortcite{car-etal:evolution,mcc-wei:anisotropic}}.  Coherent
vortices in the ocean with spatial scales of tens of kilometers can
live for more than a year and travel across the ocean, affecting the
energetics, salinity, and biology of the ocean. The number and
strength of such vortices is often determined by manually identifying
vortices. One method of validating atmospheric models is determining
whether they capture the statistics of atmospheric vortices such as
storms and hurricanes.  In all these areas, a robust efficient method
to calculate the vortex statistics would represent a major advance.

\begin{figure}[!t]
\begin{center}
\includegraphics*[width=5in]{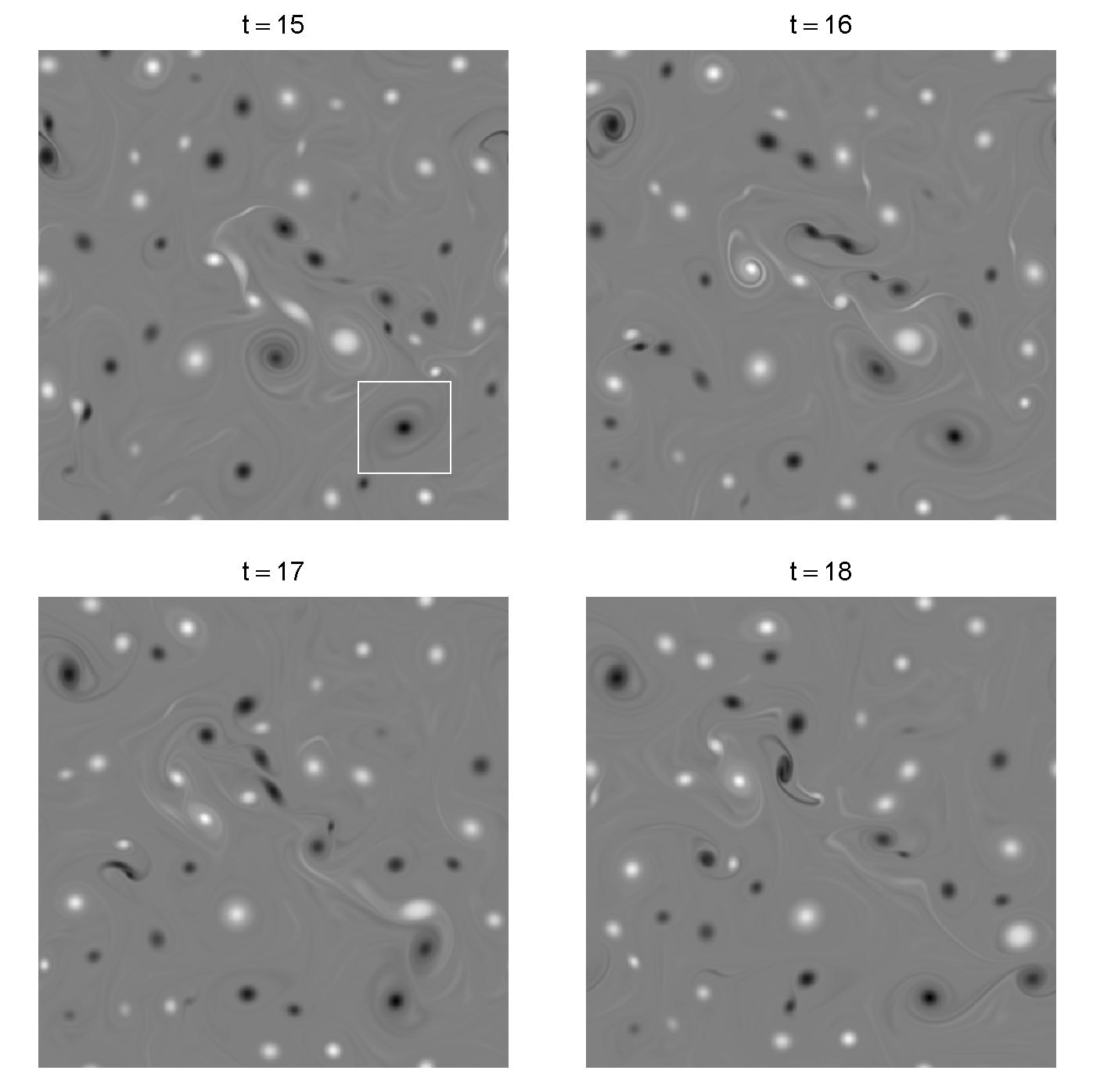}
\caption{Observed vorticity fields ($512{\times}512$ pixels) from a
  numerical simulation at times $t\in\{15,16,17,18\}~\text{seconds}$.
  Negative vorticity is seen as darker shades of grey approaching
  black, while positive vorticity is seen as lighter shades of grey
  approaching white.  The background, roughly zero vorticity, is a
  medium grey.  The white square in the lower-right corner of the
  vorticity field at $t=15$ is centered at the largest coherent
  structure, in peak absolute vorticity.}
\label{fig:vortex-field}
\end{center}
\end{figure}

% In the atmosphere and ocean, turbulence is an important physical
% feature moderating how fluids mix and move.
Figure~\ref{fig:vortex-field}, which displays observations from a
numerical simulation of turbulent flow, provides examples of such
vortices.  A vortex is a spinning, turbulent flow that possesses
anomalously high (in absolute value) vorticity.  Following the
definition common for the Norther Hemisphere, positive vorticity
(lighter shades in the images) corresponds to spinning in the
counter-clockwise direction and negative vorticity (darker shades in
the images) corresponds to spinning in a clockwise direction.  Regions
of high vorticity exhibit a peak near its center and decays back to
zero vorticity in all directions from that center.  The
self-organization of a fluid into vortices is an emergent phenomena
that can only be partially understood by analysis of the governing
partial differential equations.  Determining the details of a vortex
population requires analyzing a time-evolved field, either from
numerical simulations, laboratory experiments, or observations of
natural systems, and requires a pattern recognition (or census)
algorithm.
% Although vortices have an impact on fluid flow, their properties
% cannot be deduced directly from the equations of motion used to
% describe turbulence, but instead must currently be quantified via
% numerical simulation.
% {\bf Point (c) of referee 1: he says that this sentence is
% confusing...}

% Indeed, the richness of coherent structures in fluid flow is not
% obvious from a perusal of the governing physical equations.

In order to determine the statistics of the vortices, one must first
identify individual vortices and measure their properties.  In
two-dimensional turbulence the structure of the vortices is relatively
simple and a broad variety of census algorithms have been successful
{\cite{mcw:vortices,far-phi:coherent,sie-wei:census}}.  However one
would like to develop methods of structure identification that work in
more realistic fluid situations ranging from three-dimensional
idealized planetary turbulence to the most realistic General
Circulation Model (GCM).  In these situations, the structures include
jets as well as vortices, and these structures exist in a more complex
fluctuating environment.  The goal of the current work is to develop a
census algorithm for identifying vortices in two-dimensional
turbulence that is sufficiently general to handle, with modifications,
these more realistic situations.

%%%%%%%%%%%%%%%%%%%%%%%%%%%%%%%%%%%%%%%%%%%%%%%%%%%%%%%%%%%%%%%%%%%%%%%%%
%%%%%%%%%%%%%%%%%%%%%%%%%%%%%%%%%%%%%%%%%%%%%%%%%%%%%%%%%%%%%%%%%%%%%%%%%

The ``data mining'' of turbulent fluid flow, both simulated and
observed, can be thought of as a statistical problem of feature
extraction.  In this work we focus on the detection of coherent
structures from a simulated scalar field of rotational motion.  The
reason that we focus on simulated fields is that real data at the
required resolution is difficult to obtain in practice.  Our approach
to this problem consists of two parts.  The first step is to develop a
flexible model for a single coherent ``template'' structure (vortex)
using multiresolution analysis and the second identifies individual
vortices through a stepwise model selection procedure.  Although the
model for the template function embodies prior information from the
scientist, it is flexible enough to capture a broad range of features
associated with coherent structures one might observe in fluid.  Once
a suitable template is chosen, the second step provides an objective
approach to identifying features of interest through classic
statistical methodology, allowing the information in the vorticity
field to dictate what is and what is not a vortex.  With individual
coherent structures efficiently summarized through a set of
parameters, such as diameters and centroid locations, the next step of
modeling movement and vortex interaction may take place; e.g.,
{\shortciteN{sto-etal:asilomar}}.  This has the potential to advance
our understanding of many complex natural systems, such as hurricane
formation and storm evolution.

In the past, coherent structures have been quantified by simply
decomposing the flow into signal and background noise using
wavelet-based techniques; see, for example,
{\shortciteN{far-etal:improved}}, {\shortciteN{wic-etal:ecwpalc}},
{\shortciteN{wic-far-goi:tdcst}} and the summary article by
{\shortciteN{far-etal:turbulence}}.  The focus of the methodology used
in these papers has a engineering orientation; that is, the wavelet
transform is used to discriminate between signal and noise using
compression rate as a classifier.  The idea being that the coherent
structures (vortices) observed in the turbulent flow will be
well-represented by only a few wavelet coefficients and therefore most
coefficients may be discarded.  From one vorticity field, two fields
are produced from such a procedure: one based on the largest wavelet
coefficients that is meant to capture the largest coherent structures
in the original field; and another, based on the remaining wavelet
coefficients, is a mixture of background structures (i.e., filaments)
and noise.

{\citeN{sie-wei:census}} developed a wavelet-based census algorithm
for two-dimensional turbulence that relied on specific attributes of
coherent structures.  First, the vorticity field was separated into
coherent and background fields using an iterative wavelet thresholding
technique, then the surviving wavelet coefficients in the coherent
part are grouped according to the spatial support of the Haar basis.
{\citeN{luo-lel:eddies}} investigated a wavelet-based technique for
identifying, labeling and tracking ocean vortices and a database was
built of wavelet signatures based on the amount of energy contained at
each scale.  They found that Gaussian densities were adequate to model
observed vortices.

Our methodology differs significantly from the signal-plus-noise model
and other wavelet-based techniques.  Our main scientific contribution
is the development and efficient implementation of a statistical
method to extract a wide variety of coherent structures from
two-dimensional turbulent fluid flow.  We achieve this through the
formulation of a flexible statistical model for individual coherent
structures (vortices) and obtain a fixed number of isolated vortices
from the original field via regression methodology.  From these models
a completely different set of summary statistics may be calculated;
for example, instead of global summary statistics like the enstrophy
spectrum over the entire image we are able to look at local statistics
for each structure such as the average size, amplitude, circulation
and enstrophy of the individual vortices.

This work has lead to new models for two-dimensional multiscale
features.  Although this work concentrates on vortices, the proposed
methodology is quite generic and can be transferred to model other
types of coherent structures.  The adaptation of model selection
methods for segmenting images is made possible by formulating our
model in terms of multiple linear regression.  Finally, we note this
work also depends on the development of computing algorithms whereby
one fits many local regressions simultaneously using equivalences with
image convolution filtering.  The statistical computing aspects of
this problem are important in order to analyze large problems that are
of scientific interest.  As a test of these ideas we are able to carry
out a census of vortices from a high-resolution simulation of
two-dimensional turbulence.

The next section provides a brief introduction to two-dimensional
turbulence and the numerical simulation used.
Section~\ref{sec:models} outlines how we model individual coherent
structures and the entire vorticity field.  The two-dimensional
multiresolution analysis approach to images is introduced and the
user-defined vortex template function is also discussed.
Section~\ref{sec:estimation} provides the methodology to estimate a
single coherent structure and also multiple structures in the
vorticity field.  Section~\ref{sec:application} looks at how well the
technique performs at estimating multiple coherent structures and at
the temporal scaling of vortex statistics.

\section{Two-dimensional Fluid Turbulence}

Turbulence in atmospheres and oceans occur in many forms and at many
scales.  Two-dimensional turbulent flow is an idealization that
captures many of the features of planetary turbulence, and in
particular, has vortex behavior that is common to many atmospheric and
oceanic phenomena.  Thus, two-dimensional turbulence often serves as a
laboratory to study aspects of planetary turbulence and as a testbed
for developing theories and algorithms.  Here we follow this strategy
and use two-dimensional turbulence to develop and test a new
multiresolution census algorithm for vortex statistics.  While there
are experimental fluid systems that approximate two-dimensional fluid
flow, numerical simulations are the most effective way to develop and
test such algorithms.  Treating the output of numerical simulations as
data which is analyzed just as one would with observations or
laboratory experiments is routine in the field of turbulence and we
follow this route.

Two dimensional fluid dynamics has a long history and many aspects are
well understood
{\cite{kra-mon:turbulence,fri:turbulence,les:turbulence}}.  The
equations of motion for two-dimensional turbulence are written in
terms of the fluid velocity $\vec{u}=(u_x,u_y)$, and its scalar
vorticity
\[
\zeta(\vec{u}) = (\nabla \times \vec{u}) \cdot \hat{z},
\]
where $\nabla$ is the gradient operator and $\hat{z}$ is a unit vector
in the direction perpendicular to the plane containing the
two-dimensional velocity $\vec{u}$.  In general the curl of a vector
field is given by $\nabla\times\vec{u}$ and, hence, vorticity is the
curl of the velocity field.  Following the ``right-hand rule''
vorticity is positive when the flow is rotating anti-clockwise and
negative otherwise.  A related concept is circulation which is related
to vorticity by Stoke's theorem
\[
\Gamma(\vec{u}) = \int_S \zeta(\vec{u}) \,\text{d}S,
\]
where $S$ is a surface in two dimensions.  The units of circulation
are length squared over time and vorticity is the circulation per unit
area.  Enstrophy is given by 
\[
\mcal{E}(\vec{u}) = \frac{1}{A} \int_S |\zeta(\vec{u})|^2 \,\text{d}S,
\]
where $A$ is the area of the fluid, and is a measure of the
mean-square vorticity.  Enstrophy is a quantity that is similar to
energy (which is the mean-square velocity) and plays an important role
in turbulence theory despite being somewhat non-intuitive. 
% and is related to the kinetic energy in the flow associated with the
% dispersion effects in the fluid.  Given the basic scientific goal is
% to understand how the number and size of vortices can change over
% time and also how they interact, the statistical challenge is to
% estimate quantities of interest (e.g., circulation, enstrophy, etc.)
% from these coherent structures from a sequence of images in time --
% this is defined to be a census.  {\bf Point (d) of referee 1: what
% is the value of knowing the circulation and enstrophy?}

It is important to understand fluid flow at a macroscopic scale and,
in particular, create more accurate models for the flow of the
atmosphere and ocean.  Recent advances in computing have allowed
scientists to produce more realistic simulations of turbulent flows
{\cite{fer:eddy,moi-mah:DNS}}.  Numerical simulation of
two-dimensional turbulence show that random initial conditions will
self-organize into a collection of coherent vortices which
subsequently dominate the dynamics
(McWilliams~\citeyearNP{mcw:emergence,mcw:vortices}).  Due to this
self-organization, traditional scaling theories of turbulence fail to
correctly describe the dynamics, while scaling theories based on the
statistics of the vortices are much more successful
{\shortcite{car-etal:evolution,wei-mcw:temporal-scaling,bra-etal:revisiting}.

% With the current array of numeric simulations comes the opportunity
% to identify individual structures and better understand their
% attributes. The problem is that coherent structures must be
% discerned from the fluid flow.  This is a difficult image analysis
% problem because structures such as vortices have features at several
% spatial scales and vary in shape.  Given the basic scientific goal
% is to understand how the number and size of vortices can change over
% time and also how they interact, the statistical challenge is to
% estimate the statistics of these coherent structures from a sequence
% of images in time -- this is defined to be a census.  Previous work
% on vortex census algorithms has improved scientist's understanding
% of turbulence and led to improved theoretical models, but has been
% limited to very simplified simulations.  The statistical procedure
% presented here promises to advance the study of coherent structures
% from relatively simple simulations to more complicated models (such
% as output from a general circulation model, GCM) and observed
% satellite imagery.

\subsection{Data description}

The equations of motion for two-dimensional fluid flow are
\[
\frac{\partial \zeta}{\partial t} + (\vec{u} \cdot \nabla) \zeta =
     \mcal{D} \quad \text{and} \quad \zeta = (\nabla\times\vec{u})
     \cdot \hat{z},
\]
where $\mcal{D}$ is a general dissipation operator.  Despite being
small, dissipation is an important component of turbulence and is
necessary for numerical simulations.  Due to the nature of
two-dimensional turbulence, energy cascades to large scales and is not
dissipated in the limit of small dissipation, while enstrophy cascades
to small scales and has finite dissipation in the limit of small
dissipation.  The dissipation of enstrophy is governed by the
evolution of the vortex population.  These equations are simulated
with doubly periodic boundary conditions using a pseudo-spectral
algorithm, using a hyperviscous diffusion operator
$\mcal{D}=-\nu\nabla^4\zeta$ on a $512{\times}512$ grid.  The
simulations start from small-scale random initial conditions.  
% {\bf Point (e) of referee 1: why is dissipation included in the
% definition?}

As time proceeds, the random initial conditions self-organize into
coherent vortices.  The initial scale of the vortices is governed by
the scale of the initial conditions.  Subsequently, vortices grow
through vortex mergers until a final end-state is reached with two
vortices, one of each sign.  During the time period after
self-organization but before the vortex number gets too small, the
turbulence can be modeled as a population of interacting coherent
vortices.  We emphasize that the vortices are not introduced into the
flow by any external forcing, but rather they arise through the
natural self-organizing nature of two-dimensional turbulence.

% The vortex fields generated by the simulation used here are
% representative of populations of coherent vortices arising in other
% simulations as well as in laboratory experiments and in atmospheres
% and oceans.  Thus, the results of this work are broadly
% applicable. {\bf From AE: Please explain more why we only deal with
% simulated fields.}

Here we apply our algorithm solely to data from numerical experiments.
% While such an approach may be unusual in the statistics community,
% it is quite common in fluid dynamics.  
The equations of motion and the methods for solving them are firmly
established and previous work in fluid dynamics has shown that
numerical simulations, laboratory experiments, and observations of
natural systems all produce vortices with similar properties.
Numerical simulations provide the most complete and accurate
representation of vortex dynamics and thus provide the most stringent
test for the proposed algorithm. 

\section{Models for Vorticity Fields}
\label{sec:models}

The goal of our multiresolution census algorithm is to identify all
coherent structures (vortices) from a given vorticity field and then
compute summary statistics from each identified structure.  As
discussed in the introduction, this approach differs from previous
ones in that we are not interested in merely separating the coherent
portion of the field from the background, as in the traditional
signal-plus-noise model, but instead isolate individual coherent
structures in the image for further analysis.  For the simple vortex
fields considered here, our proposed method performs similarly to
previous methods.  This is quantitatively demonstrated by the scaling
relations shown in Figure~\ref{fig:scaling}.  However, previous
methods are not easily extended to flows with more complex structure.
The general statistical framework of the current approach provides a
flexible modelling framework in which to implement generalizations in
a relatively straightforward manner.

% The goal of our multiresolution census algorithm is to identify all
% coherent structures (vortices) from a given vorticity field and then
% compute summary statistics from each identified structure.  As
% discussed in the introduction, this approach differs from previous
% ones in that we are not interested in merely separating the coherent
% portion of the field from the background, as in the traditional
% signal-plus-noise model, but instead isolate individual coherent
% structures in the image for further analysis.

For clarity we first outline a continuous version of the problem and
then follow by a more practical discrete approximation.  Let
$\zeta_t(\bld{x})$ denote an observed vorticity field at time $t$ and
location $\bld{x}\in\bb{R}^2$.  Given $S$ vortices, $\zeta_t(\bld{x})$
can be decomposed as
\begin{equation}\label{eqn:vortex+background}
  \zeta_t(\bld{x}) = \sum_{s=1}^S v_s(\bld{x}) + e_t(\bld{x})
\end{equation}
where $v_s$ is the localized vorticity associated with each coherent
structure and $e_t$ is the background variation from other types of
structures.  Given the dominance of vortices in our application, it is
useful to assume a stochastic character for the residual component
$e_t$.  The statistical problem is: given the observed field
$\zeta_t$, estimate $S$ and $\{v_s\}$.

The main modeling component in this work is to expand the individual 
vortex field in a finite basis
\begin{equation}\label{eqn:individual}
  v_s = \sum_{i=1}^M \alpha_{s,i} z_i(\bld{x} - \bs{\mu}_s)
\end{equation}
where $\bs{\mu}_s$ approximates the center of the vortex and the
(linear) coefficients $\{\alpha_{s,i}\}$ determine the shape.
Estimation procedures for $\bs{\mu}_s$ and $\{\alpha_{s,i}\}$ are
discussed in Section~\ref{sec:estimation}.  The basis functions
$\{z_i\}$ are designed to provide a multiscale representation of
coherent structures and build in prior knowledge of the vortex shape.
Their specification is motivated in the next two sections.

\subsection{Multiresolution Analysis (MRA)}
\label{sec:MRA}

The use of wavelets for two-dimensional image analysis and compression
has a large literature and at its heart is the decomposition of an
image into different scales or levels of resolution
{\cite{vet-kov:wavelets,mal:tour}}.  In our work we assume the scale
to be in powers of two and within a scale consider a further
decomposition that divides features into three orientations: vertical,
horizontal, and diagonal.  The net result is that for a given initial
vorticity field $\zeta_t$ and $J$ levels of resolution ($J=6$ is used
when applying the model in practice), we decompose the field into the
sum of $(3J+1)$ distinct components: three different orientations for
each level of resolution and a smoothed field.  Heuristically, as the
levels increase the main features at each level will increase in size
by a factor of two.  Strong horizontal features at a given scale are
represented by the horizontal component while similar correspondences
hold for the vertical and diagonal components.  Later, it will be
exploited that some of these level-specific features may correspond to
features at (possibly different) levels of the decomposition of the
vorticity field.

% The key idea in selecting a template function is to use the wavelet
% basis functions derived from decomposing a simple representation of
% a vortex.  It remains to develop a rigorous mathematical description
% of this process and a computational algorithm to produce the
% decomposition.

We assume that the vorticity field $\zeta(\bld{x})$ has finite energy;
i.e., $\int_\bld{x}\zeta^2(\bld{x})<\infty$ and let $\phi$ be a
scaling function and $\psi$ be the corresponding wavelet generating 
an orthonormal basis on $L^2(\bb{R})$.  Define the three separable
two-dimensional wavelets
\[
\psi^{\text{h}}(x_1,x_2) = \phi(x_1)\psi(x_2), \quad
\psi^{\text{v}}(x_1,x_2) = \psi(x_1)\phi(x_2), \quad
\psi^{\text{d}}(x_1,x_2) = \psi(x_1)\psi(x_2),
\]
corresponding to the horizontal, vertical, and diagonal directions,
respectively.  This follows from the fact that the two-dimensional
wavelets are the outer product of two one-dimensional wavelet and
scaling functions, where the scaling function averages (smooths)
across its spatial direction while the wavelet function differences
across its spatial direction.  The two-dimensional wavelet
$\psi^\text{h}(x_1,x_2)=\phi(x_1)\psi(x_2)$ will therefore smooth
across the first dimension ($x_1$-axis) and difference across the
second dimension ($x_2$-axis), thus favoring horizontal features.  The
two-dimensional wavelet basis function
$\psi^\text{v}(x_1,x_2)=\psi(x_1)\phi(x_2)$ differences across the
$x_1$-axis and smooths across the $x_2$-axis, thus favoring vertical
features and $\psi^\text{d}(x_1,x_2)=\psi(x_1)\psi(x_2)$ differences
across both directions and favors non-vertical/non-horizontal (i.e.,
diagonal) features.  The separable scaling function
$\phi(x_1,x_2)=\phi(x_1)\phi(x_2)$ is associated with the
approximation space.  

Although the 2D~DWT, as a decimated orthonormal transform, would be a
more efficient representation of the vorticity field we find it
advantageous to utilize the 2D maximal overlap DWT (2D~MODWT).  Unlike
the orthonormal transform, the 2D~MODWT produces a redundant
non-orthogonal transform.  The reason for this discrepancy is that the
2D~MODWT does not subsample in either dimension, it only filters the
original image.  The advantages are that the transform is translation
invariant to integer shifts in space and it reduces potential
artifacts caused by how the wavelet filter represents abrupt changes
in the image.

Assume both $\psi$ and $\phi$ have been rescaled so that the squared
norm of the wavelet coefficients equals the squared norm of the
original observations.  The separable wavelet functions associated
with specific scale and spatial directions
$D\in\{\text{h},\text{v},\text{d}\}$ are given by
\[
\psi_{j,k,l}^D(x_1,x_2) = \frac{1}{2^{2j}}
\psi^D\left(\frac{x_1-k}{2^j}, \frac{x_2-l}{2^j}\right),
\quad j=1,\ldots, J, \quad k=1,\ldots,M, \quad l=1,\ldots,N.
\]
The separable scaling function $\phi_{j,k,l}$ is defined similarly.
Hence, each level in the transform will have the same spatial
dimension as the original field $(M\times{N})$ and represent a
redundant set of wavelet coefficients.  The 2D~MODWT begins with the
original vorticity field $\zeta$ (the spatial location will be omitted
when implied), and at all scales we denote
$\omega_j^\phi=\langle{\zeta,\phi_{j,k,l}}\rangle$ and
$\omega_j^D=\langle{\zeta,\psi_{j,k,l}^D}\rangle$ for
$D\in\{\text{h},\text{v},\text{d}\}$, where $\langle{x,y}\rangle$ is
the two-dimensional inner product.  The vorticity field
$\zeta(\bld{x})$ may now be decomposed into $3J+1$ sub-fields: three 
fields of wavelet coefficients at each resolution level corresponding
to distinct spatial 
directions and one field containing the scaling coefficients at the
final level.  The scaling (approximation) field for level~$j$ may be
obtained from the four fields at level~$j+1$ via
\begin{equation}\label{eqn:reconstruction}
  \omega_j^\phi(\bld{x}) = \left\{\omega_{j+1}^\phi *
    \phi_{j,k,l}\right\} + \sum_D \left\{\omega_{j+1}^D *
    \psi_{j,k,l}^D\right\},
\end{equation}
where ``$*$'' denotes the convolution operator.

The two-dimensional multiresolution analysis} (MRA) of the vorticity
field is an additive decomposition given by recursively applying
(\ref{eqn:reconstruction}) over all~$j$; i.e.,
\[
\zeta(\bld{x}) = \left\{\omega_J^\phi * \phi_{J,k,l}\right\} +
\sum_{j=1}^J \sum_D \left\{\omega_j^D * \psi_{j,k,l}^D\right\} =
\alpha_J^\phi(\bld{x}) + \sum_{j=1}^J \sum_D \alpha_j^D(\bld{x}),
\]
where $\alpha_J^\phi$ is the wavelet approximation field and
$\alpha_j^D$ is the wavelet detail field associated with the spatial
direction $D\in\{\text{h},\text{v},\text{d}\}$.  The MRA of
$\zeta(\bld{x})$ provides a convenient way of isolating features at
different scales and directions with coefficients in the spatial
domain versus the wavelet domain.  This is advantageous since
reconstruction is now reduced from the full inverse 2D~MODWT to simple
addition and potential phase adjustments are eliminated.

\begin{figure}[!t]
\begin{center}
\includegraphics*[width=.75\textwidth]{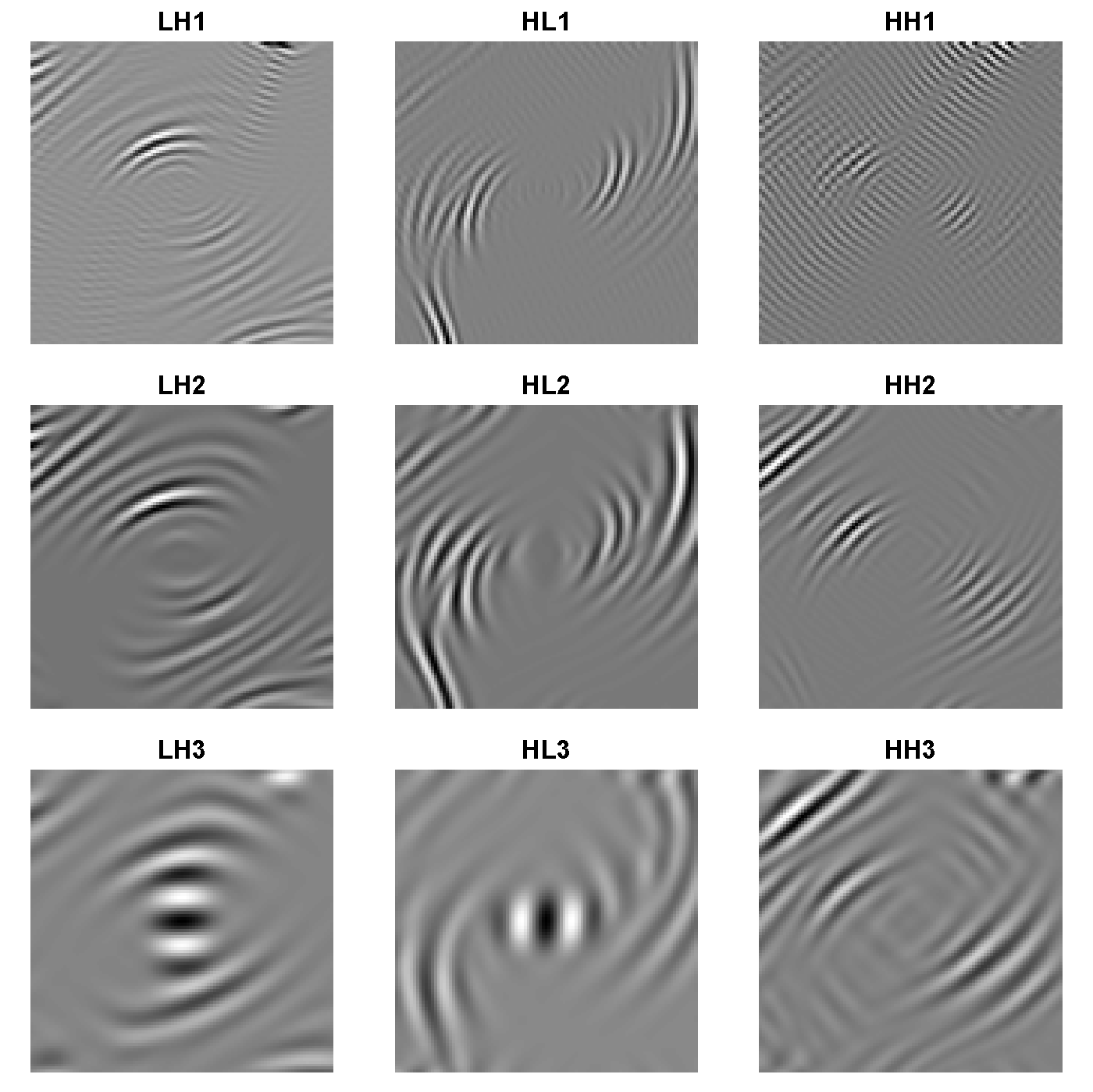} 
\caption{Two-dimensional multiresolution analysis $(j=1,2,3)$ of a
  $100{\times}100$ pixel section of the vorticity field centered at
  $(x,y)=(399,101)$ using the Daubechies least asymmetric wavelet
  filter $(L=8)$.  LH corresponds with horizontal, HL with vertical
  and HH with diagonal wavelet coefficients.  This pixel section is
  the area highlighted by the white square box in the top left panel
  of Figure~\protect\ref{fig:vortex-field}.}
\label{fig:vort1}
\end{center}
\end{figure}
%%%%%%%%%%%%%%%%%%%%%%%%%%%%%%%%%%%%%%%%%%%%%%%%%%%%%%%%%%%%%%%%%%%%%%%%%

%%%%%%%%%%%%%%%%%%%%%%%%%%%%%%%%%%%%%%%%%%%%%%%%%%%%%%%%%%%%%%%%%%%%%%%%%
\begin{figure}[!t]
\begin{center}
\includegraphics*[width=.75\textwidth]{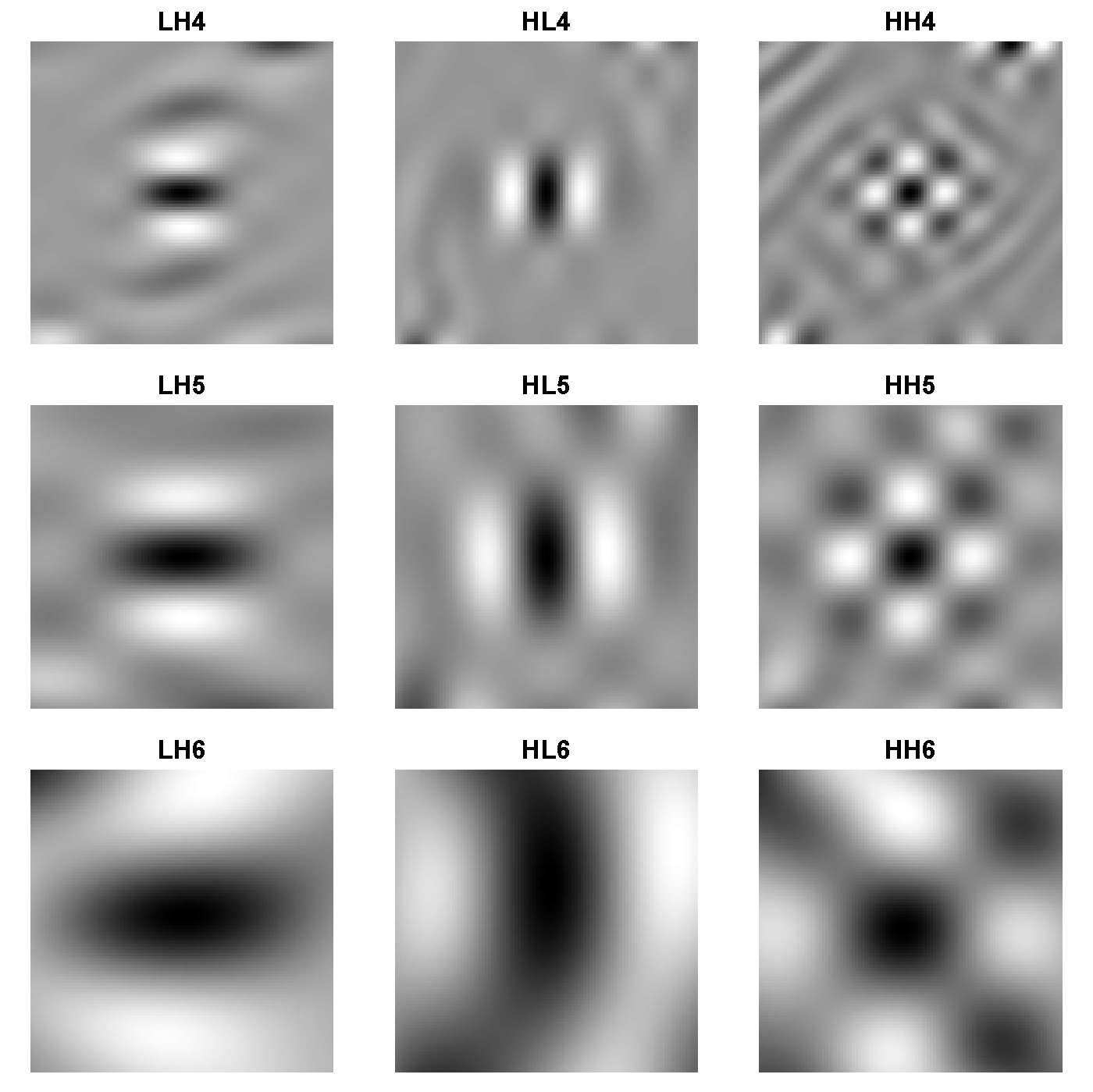} 
\caption{Two-dimensional multiresolution analysis $(j=4,5,6)$ of a
  $100{\times}100$ pixel section of the vorticity field centered at
  $(x,y)=(399,101)$ using the Daubechies least asymmetric wavelet
  filter $(L=8)$.  LH corresponds with horizontal, HL with vertical
  and HH with diagonal wavelet coefficients.}
\label{fig:vort2}
\end{center}
\end{figure}
%%%%%%%%%%%%%%%%%%%%%%%%%%%%%%%%%%%%%%%%%%%%%%%%%%%%%%%%%%%%%%%%%%%%%%%%%

Figures~\ref{fig:vort1} and~\ref{fig:vort2} display the six scales
from a two-dimensional MRA of the sample vorticity field in
Figure~\ref{fig:vortex-field} (at $t=15$), defined by a
$100{\times}100$~pixel section centered at $(x,y)=(399,101)$.  Each
row displays the wavelet detail fields associated with the three
spatial directions: horizontal, vertical and diagonal.  It is clear
that each of the two-dimensional wavelet filters captures distinct
spatial directions at a fixed spatial scale.  Given the filaments from
this particular vortex are elliptical in shape, it is not surprising
to see the detail coefficients of the filament structures strongest in
the northeast-southwest directions.  It is interesting to note that
the coherent structure (i.e., the dark region in the center of the
image) is not seen until the third scale (third row in
Figure~\ref{fig:vort1}) and then only in the horizontal and vertical
directions.  At higher scales, corresponding to larger spatial areas
and lower spatial frequencies, the coherent structure is apparent in
all three directions.  This is most likely due to the spatial extent
(size) of the structure at time $t=15$ in the simulation.

\subsection{Single Vortex Model}
\label{sec:single-vortex}

\begin{figure}[!t]
\begin{center}
\includegraphics*[width=.95\textwidth]{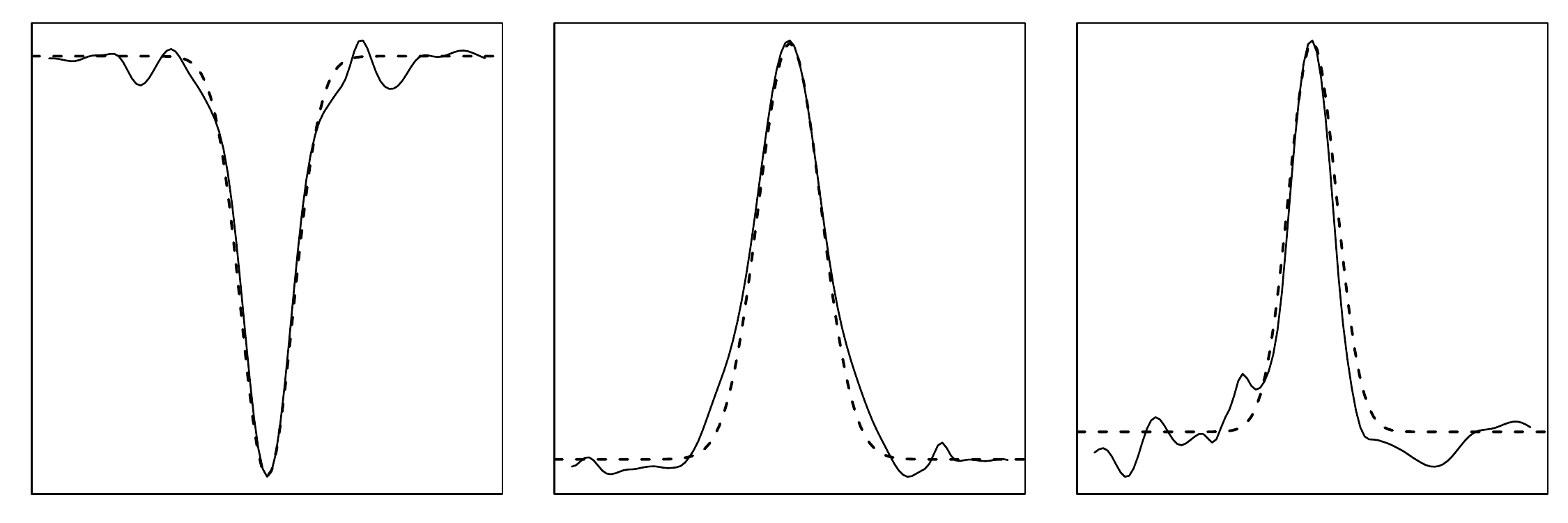} 
\caption{Vortex cross sections (solid line) and normalized Gaussian
  kernels (dashed line) from the observed vorticity field at time
  $t=15$.}
\label{fig:cross-section}
\end{center}
\end{figure}

Recall that a vortex may be loosely described as a concentration of
anomalously high (in absolute value) vorticity.  As a simple outline
of a vortex we consider the Gaussian kernel.  If we translate a single
vortex to the origin, then
$\tau(\bld{x})=\eta\exp(-\norm{\bld{x}}^2/\sigma^2)$ is our vortex
template function where $\eta$ is the maximum vorticity at its center.
This choice for $\tau(\bld{x})$ visually appears to capture the
relevant features of an idealized vortex even when the observed
vortices decay back to zero at a different rate than Gaussian tails.
For illustration in Figure~\ref{fig:cross-section}, $\eta$ and
$\sigma^2$ were chosen to coincide with the peak value and spread of
the specific coherent structure.  However, when used in the
multiresolution census algorithm the template function will have a
fixed magnitude and spread -- any modifications to fit the vortex will
be induced by the multiscale representation of $\tau(\bld{x})$.  Thus,
a perfect fit is not required since deviations from Gaussianity will
be captured through the model fitting procedure.

%%%%%%%%%%%%%%%%%%%%%%%%%%%%%%%%%%%%%%%%%%%%%%%%%%%%%%%%%%%%%%%%%%%%%%%%%
%%%%%%%%%%%%%%%%%%%%%%%%%%%%%%%%%%%%%%%%%%%%%%%%%%%%%%%%%%%%%%%%%%%%%%%%%

%%%%%%%%%%%%%%%%%%%%%%%%%%%%%%%%%%%%%%%%%%%%%%%%%%%%%%%%%%%%%%%%%%%%%%%%%
\begin{figure}[!t]
\begin{center}
\includegraphics*[width=.75\textwidth]{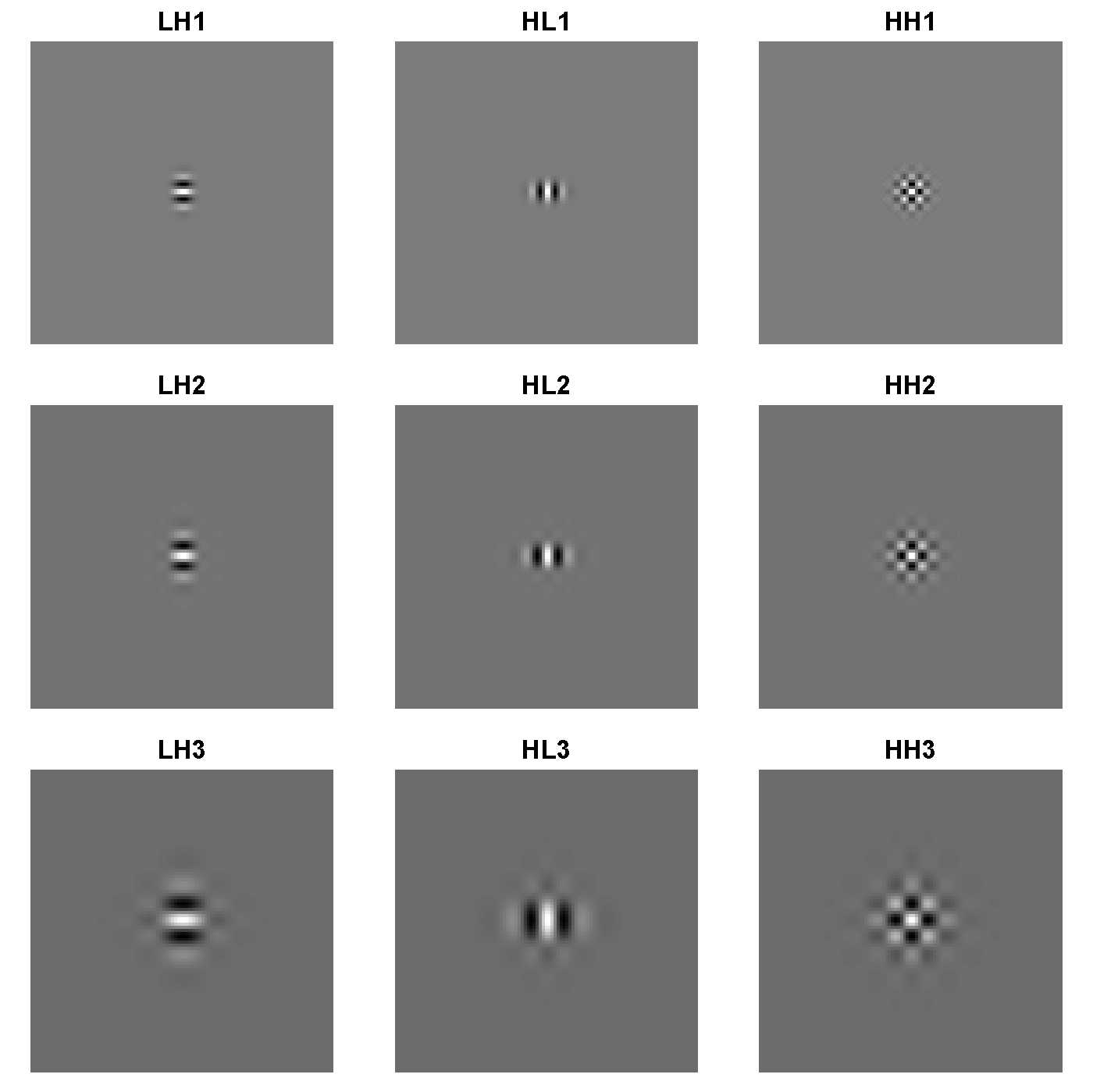} 
\caption{Two-dimensional multiresolution analysis $(j=1,2,3)$ of a
  $100{\times}100$~pixel section of the template function.  The rows
  correspond to the scale $j$ and the columns correspond to spatial
  directions: LH is horizontal, HL is vertical and HH is diagonal.}
\label{fig:temp1}
\end{center}
\end{figure}
%%%%%%%%%%%%%%%%%%%%%%%%%%%%%%%%%%%%%%%%%%%%%%%%%%%%%%%%%%%%%%%%%%%%%%%%%

%%%%%%%%%%%%%%%%%%%%%%%%%%%%%%%%%%%%%%%%%%%%%%%%%%%%%%%%%%%%%%%%%%%%%%%%%
\begin{figure}[!t]
\begin{center}
\includegraphics*[width=.75\textwidth]{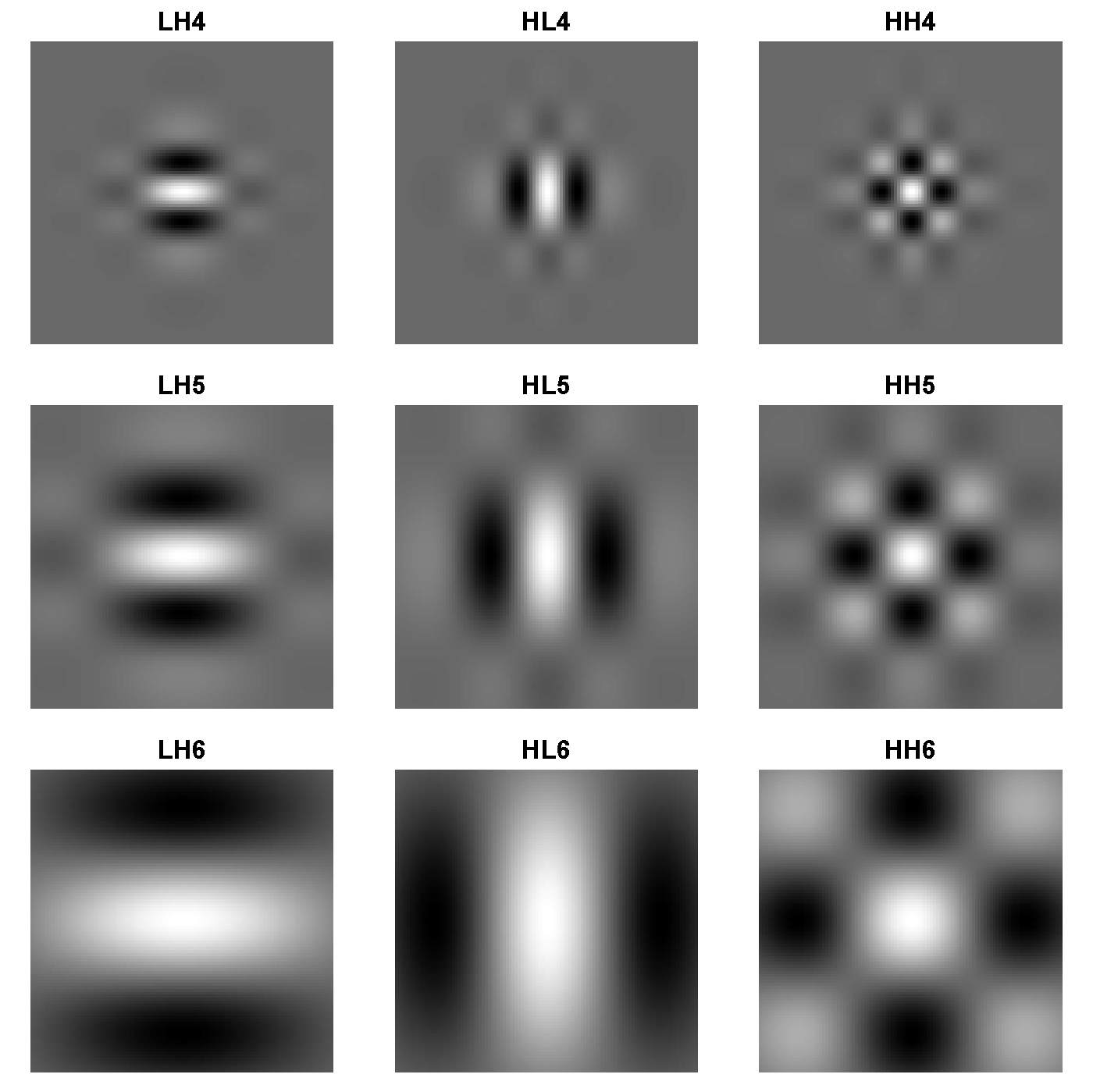}
\caption{Two-dimensional multiresolution analysis $(j=4,5,6)$ of a
  $100{\times}100$~pixel section of the template function. The rows
  correspond to the scale $j$ and the columns correspond to spatial
  directions: LH is horizontal, HL is vertical and HH is diagonal.}
\label{fig:temp2}
\end{center}
\end{figure}
%%%%%%%%%%%%%%%%%%%%%%%%%%%%%%%%%%%%%%%%%%%%%%%%%%%%%%%%%%%%%%%%%%%%%%%%%

Figures~\ref{fig:temp1} and~\ref{fig:temp2} show an MRA $(J=6)$ of
$\tau(\bld{x})$ derived from the outer product of two zero-mean
Gaussian kernels with the same variance.  The rows correspond to wavelet
scales and the columns correspond to spatial directions within each
scale.  Although the Gaussian kernel is too simple for modeling an
observed vortex directly, the components of its MRA are strikingly
similar to the MRA of individual coherent structures in the observed
vorticity field (Figures~\ref{fig:vort1} and~\ref{fig:vort2}).  This
observation identifies the basic components for building a template
basis and is key to our methodology.  We propose to represent each
matrix of vorticity coefficients as a linear combination of the
matrices derived from the MRA of the Gaussian kernel.  A straightforward
way to relate the MRA of an observed coherent structure to the
template function is through a series of simple linear regression
models relating the multiresolution coefficients from the observed
vorticity field $\bs{\alpha}$ with the multiresolution coefficients
from the template function $\bld{z}$ via
\begin{equation}\label{eqn:simple}
  \begin{array}{lcll}
    \alpha_{j,k,l}^D & = & \beta_j^D z_{j,k,l}^D + \epsilon_{j,k,l}^D,
    & \quad k=1, \ldots, M, \quad l = 1, \ldots, N; ~ \text{for all $j,D$}\\ 
    \alpha_{J,k,l}^{\phi} & = & \xi_J^{\phi} + \beta_J^{\phi}
    z_{J,k,l}^{\phi} + \epsilon_{J,k,l}^{\phi}, & \quad k=1, \ldots,
    M, \quad l = 1, \ldots, N.
  \end{array}
\end{equation}
At each spatial scale and direction there is only one final parameter
to be fit, since each image from the MRA is guaranteed to be mean zero
except for the smoothed field $\bs{\alpha}^{\phi}$.  The implied
linear relation between the multiresolution components of the data and
template function allows for differences in the magnitude and
direction of vorticity through the regression coefficients.  That is,
once $\tau(\bld{x})$ is defined (using fixed parameters $\eta$ and
$\sigma^2$) we will use it to model all possible coherent structures
in the vorticity field.  The MRA decomposes the template function into
spatial scales and directions, but the model in
Equation~(\ref{eqn:simple}) is limited because each spatial scale and
direction from $\zeta(\bld{x})$ is tied to the same spatial scale and
direction of $\tau(\bld{x})$.  To make full use of the MRA, we propose
to estimate every spatial scale and direction of the observed
vorticity field using all possible spatial scales and directions from
the template function.  This suggests the following multiple linear
regression model:
\begin{equation}\label{eqn:multiple}
\alpha_{j,k,l}^D = \beta_1^{\text{h}} z_{1,k,l}^{\text{h}} +
\beta_1^{\text{v}} z_{1,k,l}^{\text{v}} + \beta_1^{\text{d}}
z_{1,k,l}^{\text{d}} + \cdots + \beta_J^{\text{h}} z_{J,k,l}^{\text{h}} +
\beta_J^{\text{v}} z_{J,k,l}^{\text{v}} + \beta_J^{\text{d}}
z_{J,k,l}^{\text{d}} + \beta_J^{\phi} z_{J,k,l}^{\phi} +
\epsilon_{j,k,l}^D,
\end{equation}
$k=1,\ldots,M, l=1,\ldots,N$; for all $j,D$.  The linear regression model
for the field of wavelet smooth coefficients in
Equation~(\ref{eqn:simple}) does not change.  The intercept is
included to account for potential low-frequency oscillations that may
not be provided by the template function.  At each spatial scale and
direction from the observed vorticity field there are $(3J+1)$
parameters to be fit.  The full multivariate multiple linear
regression model for a single vortex may now be formulated as
\begin{equation}
\bld{Y} = \bld{Z} \bs{\beta} + \bs{\epsilon},
\end{equation}
where $\bld{Y}$ is the $MN{\times}(3J+1)$ response matrix from the MRA
of the observed vorticity field, $\bld{Z}$ is the $MN{\times}(3J+1)$
design matrix whose columns consist of the MRA of the template
function centered at the location $\bs{\mu}$, and $\bs{\beta}$ is the
$(3J+1)\times(3J+1)$ regression coefficient matrix a given coherent
structure.

There are several differences between the Gaussian kernel $\tau(\bld{x})$
and the observed vorticity field $\zeta(\bld{x})$ that are handled
automatically through the multivariate multiple linear regression
model in Equation~(\ref{eqn:multiple}), these include: fixed spatial
size, amplitude, and radial symmetry.  The fixed spatial size of
$\tau(\bld{x})$ is taken care of by the fact that all spatial scales
from its MRA are associated with each spatial scale in the observed
vorticity field.  The scale dependent multiple linear regression,
Equation~(\ref{eqn:multiple}), may represent any particular spatial
scale of $\zeta(\bld{x})$ and associate it with any spatial scale of
$\tau(\bld{x})$.  Thus, larger spatial scales may be either favored or
penalized through the individual regression coefficients.  The
amplitude of the template function is similarly handled through the
magnitude of these regression coefficients.

An initial impression may be that the radial symmetry of
$\tau(\bld{x})$ will restrict the model to radially symmetric
vortices.  However, asymmetry is accommodated through the multivariate
multiple linear regression by decoupling the three unique spatial
directions.  Each spatial direction is manipulated through its own
regression coefficient, thus allowing for eccentric vortex shapes by
favoring one or two of the three possible spatial directions at each
scale (Figures~\ref{fig:temp1} and~\ref{fig:temp2}).

The level of association needed between the template function and
feature of interest in order to model the vorticity field is currently
unknown.  We have found that small pixel shapes, such as a Gaussian or
triangular function, work well in the problem presented here.
However, we believe this algorithm is quite adaptive through the
user-defined template function $\tau(\bld{x})$.  For example, if the
filament structures were deemed interesting in this problem, an
effective template function to pick out these features could consist
of nothing more than a collection of concentric circles.

\section{Model Estimation}
\label{sec:estimation}

\subsection{A Linear Model Framework}
\label{sec:linear-model}

Based on the previous development of the template function and the
final step of discretization we are led to the following model for a
vortex field at a single time point:
\begin{equation}\label{eqn:multivariate}
\bld{Y} = \bld{Z}_1 \bs{\beta}_1 + \bld{Z}_2 \bs{\beta}_2 + \cdots +
\bld{Z}_S \bs{\beta}_S + \bs{\epsilon},
\end{equation}
where $\bld{Y}$ is the $MN{\times}(3J+1)$ response matrix from the MRA
of the observed vorticity field (the columns index the spatial scales
and direction where each column vector is the result of unwrapping the
$M{\times}N$ image matrix into a vector of length $MN$), $\bld{Z}_s$
is the $MN{\times}(3J+1)$ design matrix whose columns consist of the
MRA of the template function centered at the location $\bs{\mu}_s$, and
$\bs{\beta}_s$ is the $(3J+1)\times(3J+1)$ regression coefficient
matrix for the $k$th coherent structure.  Note that this model
parallels the continuous versions in
Equations~(\ref{eqn:vortex+background}) and~(\ref{eqn:individual}),
but for fixed $S$ and $\{\bs{\mu}_s\}$ is a linear model.  Given this
framework, the main statistical challenge is model selection,
determining the number of coherent structures $S$, and their locations
$\{\bs{\mu}_s\}$.

\subsection{Coherent Feature Extraction}
\label{sec:algorithm}

We choose to implement our procedure for coherent structure extraction
in a modular fashion so that we may modify specific steps without
jeopardizing the integrity of the entire method.  To this end, the
first major step is to produce a set of vortex candidate points
$\mcal{C}$ (coordinates in two dimensions); that is, a subset of all
possible spatial locations in the observed vorticity field.  Our goal
at this point is not to generate the exact locations of all coherent
structures in the field, but we want $\mcal{C}$ to contain all
possible coherent structures and additional spatial locations that are
not vortices; i.e., $MN\gg\#\mcal{C}>S$.  Although the number of
elements in $\mcal{C}$ is large, their refinement is amenable to more
conventional statistical analysis.  Vortices are selected from
$\mcal{C}$ using classic likelihood procedures to obtain a final
model.  The first step is a rough screening of the model space and
should be done with computational efficiency in mind.  The second step
devotes more computing resources on a much smaller set of models.

\subsubsection{Candidate Point Selection}

Instead of relying on specific features in the vorticity field that
are present in our current data set and may or may not be present in
future applications, we propose a model-based approach to extracting
the set of vortex candidate points $\mcal{C}$.  After selecting an
appropriate template function $\tau(\bld{x})$, we fit a single-vortex
model (Section~\ref{sec:single-vortex}) to every $M{\times}N$ spatial
locations.  This is done in a computationally efficient manner by
first performing an orthogonal decomposition on the design matrix
$\bld{Z}$ and using a discrete Fourier transform to perform the matrix
multiplications {\shortcite{whi-etal:stochastic}}.  The result is that
the multivariate multiple linear regression model in
Equation~(\ref{eqn:multivariate}) can be fit to all $M{\times}N$ grid
points in the image without the need for a massive computing
environment.  The reason for fitting the single-vortex model
everywhere is that we may now use the regression coefficient matrix
$\hat{\bs{\beta}}_s$ to indicate the presence or absence of a coherent
structure at $\bs{\mu}_s$.  This technique has the inherent flexibility of
the template function.  If a different feature was to be extracted
from the vorticity field, we would simply fit a different template
function and use its regression coefficients.

With a regression coefficient matrix for each location, a distance
measure is computed between $\hat{\bs{\beta}}$ and the identity matrix
$\bld{I}_J$, of dimension $(3J+1)$, using
\[
\Lambda = \min\left\{ \text{tr}\left[(\hat{\bs{\beta}} -
  \bld{I}_J)^2\right], \text{tr}\left[(\hat{\bs{\beta}} +
  \bld{I}_J)^2\right] \right\}.
\]
The idea is that $\Lambda$ measures the feasibility that a vortex is
centered at this location.  The value of $\Lambda$ is small when the
regression matrix $\hat{\bs{\beta}}$ is similar to the identity
matrix, and thus, the single-vortex model is similar to the template
function.  Two comparisons are made, one to the positive identity
matrix and one to the negative identity matrix, so that positive and
negative spinning vortices are favored equally.  
% ({\bf From point (f) of referee 1: how did we do the weighting?)
% The parametric image of $\Lambda$ values is smoothed, using a
% nine-point nearest-neighbor kernel with more weight on the
% orthogonal neighbors than the diagonal neighbors, and a local
% minimum search is performed on the smoothed image in order to obtain
% the set of candidate points $\mcal{C}$.
The parametric image of $\Lambda$ values is smoothed using a
nine-point nearest-neighbor kernel with weights based on Euclidean
distance.  A local minimum search is performed on the smoothed image
in order to obtain the set of candidate points $\mcal{C}$.  We found
that this technique captures all coherent structures one can identify
visually in the vorticity field along with additional features that
may or may not be vortices.

\subsubsection{Model Selection for the Vortex Field}

A final selection of coherent structures from the set of candidate
points $\mcal{C}$ is achieved through forward subset selection by
minimizing the generalized cross-validation (GCV) function
\[
\text{GCV}(S) = \frac{\text{RSS}(S)/(MN)}{[1 - p(S)/(MN)]^2},
\]
for a given choice of $S$ vortex locations.  Here $\text{RSS}(S)$ is
the residual sum of squares and $p(S)$ is the total number of
effective parameters in Equation~(\ref{eqn:multivariate}).  The
following simple method was adopted to count the total number of
effective parameters.  For each estimated $\bs{\beta}_s$ in
Equation~(\ref{eqn:multivariate}), denoted as
$\hat{\bs{\beta}}_s=(\hat{\beta}_1, \ldots,
\hat{\beta}_{(3J+1)\times(3J+1)})$, its effective number of parameters
is defined as the number of $\hat{\beta}_j$'s whose absolute value is
greater than twice the sample standard deviation of the
$\hat{\beta}_j$'s.  Then the overall total effective number of
parameters in Equation~(\ref{eqn:multivariate}) is calculated as the
sum of the effective number of parameters of $\hat{\bs{\beta}}_s$ for
all $s$.

Because of the size of this problem, it is not possible to compute
$\hat{\bs{\beta}}$ via exact linear algebra.  For an observed
vorticity field with $M=N=512$ and $J=6$, model fitting will produce
approximately 95~million regression coefficients.  Instead we use an
iterative method, {\em backfitting} {\cite{fri-stu:projection}}, to
find an approximate solution to the $S(3J+1)^2$ simultaneous linear
equations associated with Equation~(\ref{eqn:multivariate}).
Convergence for this algorithm is achieved by looking at the absolute
difference between each regression coefficient matrix
$\hat{\bs{\beta}}$ from step $i-1$ to $i$.  The number of iterations
was found to be small across a wide range of $S$, usually two to four
iterations sufficed.  We attribute the small number of iterations to
the fact that most coherent structures in $\zeta(\bld{x})$ are
spatially isolated; see, e.g., Figure~\ref{fig:vortex-field}.

\section{Application to Two-dimensional Turbulent Flows}
\label{sec:application}
Now we return to the simulated vorticity field in
Figure~\ref{fig:vortex-field} and compute the vortex statistics using 
our multiresolution census algorithm outlined in
Section~\ref{sec:algorithm}.  

\subsection{Isolated and Multiple Coherent Structures}

The vorticity field at time step $t=15$ from the numerical simulation
provides a reasonable example of multiple coherent structures embedded
within a quiescent background.  Summary statistics for this vorticity
field include an average circulation given by $\bar\zeta=384.10$, an
average enstrophy of $\bar{\mcal{E}}=397.4{\times}10^3$, and an average
maximum peak amplitude of 25.63.

\begin{figure}[!t]
\begin{center}
\includegraphics*[width=.45\textwidth]{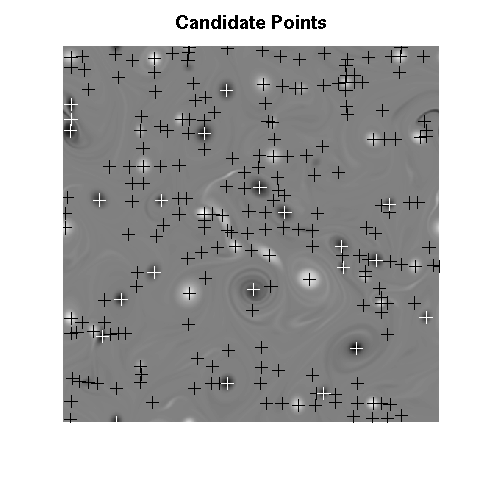}\\
\includegraphics*[width=.9\textwidth]{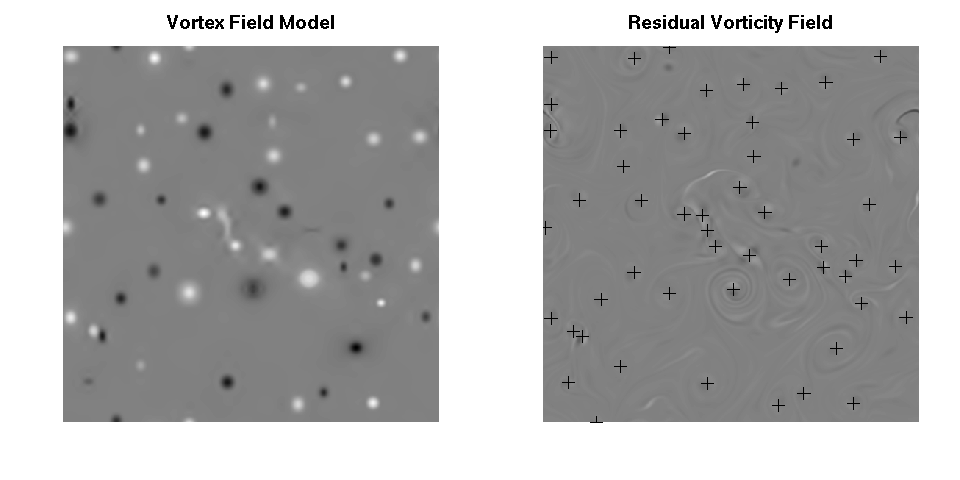}
\caption{Candidate vortex locations, the vortex field model and
  residual field for the observed vorticity field at time $t=15$,
  plotted on the same grey scale.  The spatial locations of the
  estimated coherent structures from the vortex field model are
  plotted on the residual field for comparison.  Color changes in the
  locations of the candidate points are artificial and only meant to
  visualize points in areas of large negative vorticity.}
\label{fig:fit-resid}
\end{center}
\end{figure}

To illustrate the ability of our procedure to find coherent structures
of varying sizes and shapes, Figure~\ref{fig:fit-resid} shows the
results from the two stages of our census algorithm: coherent
structures identified from our candidate procedure, the fitted vortex
field model and the residual field ($t=15$).  The set of candidate
points, roughly~200, appears to capture all potential vortices
apparent through visual inspection along with other locations that do
not appear to contain a substantial accumulation of vorticity.  After
model selection, a total number of 54~coherent structures were
identified.  Although the template function has fixed spatial size and
amplitude, the fitted coherent structures are distinct from it and
from each other, and exhibit a wide variation in both size and
amplitude.

%%%%%%%%%%%%%%%%%%%%%%%%%%%%%%%%%%%%%%%%%%%%%%%%%%%%%%%%%%%%%%%%%%%%%%%%%
%%%%%%%%%%%%%%%%%%%%%%%%%%%%%%%%%%%%%%%%%%%%%%%%%%%%%%%%%%%%%%%%%%%%%%%%%

A limitation of the current vortex field model is that elliptical
structures, such as structures being warped by others in their local
neighborhood, are not well modeled.  Although the DWT decomposes
two-dimensional structures into horizontal, vertical and diagonal
features, warped vortices do not follow both diagonal directions
simultaneously.  Hence, the diagonal elements from the template
function are not heavily utilized in the modeling process.
Alternative wavelet transforms, such as the complex wavelet transform
(Kingsbury {\citeyearNP{kin:image,kin:complex}}), produce coefficients
associated with several directions and have been more successful in
representing structures in images when compared to common orthogonal
wavelet filters.  One could also apply other types of multi-scale
transforms that are more efficient in representing and extracting
other features of interest; e.g., the curvelet transform of
{\shortciteN{sta-can-don:curvelet}}.

%Additional summary statistics from this vorticity field are an average
%circulation of 1561.4, an average enstrophy of 31956, and an average
%maximum peak amplitude of 0.64 [UNITS?].  The average maximum peak
%amplitude is provided as a fraction of the maximum peak amplitude.

\subsection{Application to Temporal Scaling}

\begin{figure}[!t]
\begin{center}
\includegraphics*[width=.75\textwidth]{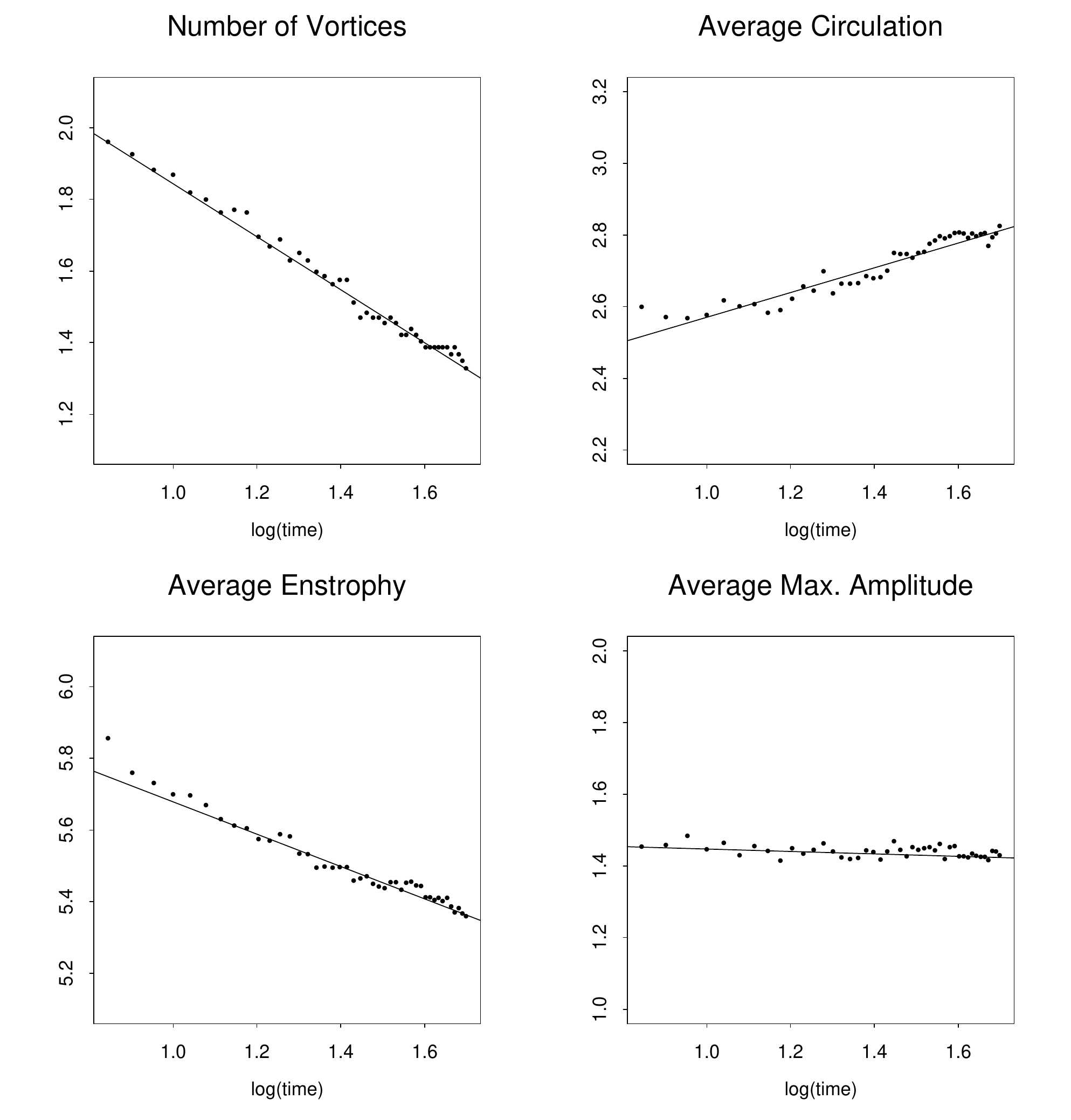}
\caption{Temporal scaling of vortex statistics from the
  multiresolution-based census algorithm.  All four quantities are
  expected to follow a scaling relation in time and are thus plotted
  on a log-log scale.  The straight line in each plot is the best
  fitting regression line.}
\label{fig:scaling}
\end{center}
\end{figure}

One scientific end point for studying a turbulence experiment is
quantifying how coherent structures and their related statistics scale
over time.  Some examples of the analysis of two-dimensional turbulent
flow are {\citeN{sie-wei:census}} using a wavelet-based procedure and
the ``objective observer'' approach of {\citeN{mcw:vortices}}.  We
applied our new method to identify the vortices in all the vorticity
fields at time $t\geq 7$.  Figure~\ref{fig:scaling} displays the
number of vortices, average circulation, average enstrophy, and
average maximum peak amplitude.

{\citeN{wei-mcw:temporal-scaling}} investigated the temporal scaling
of vortex statistics using two techniques: long-time integration of
the fluid equations and from a modified point-vortex model that
describes turbulence as a set of interacting coherent structures.  The
two systems gave the same scaling exponent $\xi\approx{0.72}$. 
% {\bf Point (g) of referee 1: what does the scaling exponent tell
% us?}
The scaling exponent measures the rate of decay of the turbulence and
is governed by the dynamics of the vortex population.  In addition, it
provides a quantitative measure to test our algorithm.
{\shortciteN{bra-etal:revisiting}} studied the evolution of vortex
statistics from very high-resolution numerical simulations.  The
results of the numerical simulation at low Reynolds number found
$\xi\approx{0.72}$, while the vortex decay rate at high Reynolds
number was $\xi\approx{0.76}$.  The Reynolds number is the ratio of
inertial forces to viscous forces.  Low Reynolds numbers correspond to
Laminar flow where viscous forces dominate and is characterized by
smooth, fluid motion and high Reynolds numbers are dominated by
inertial forces and contain vortices and jets.  The evolution of other
vortex statistics can be expressed in terms of $\xi$
{\shortcite{car-etal:evolution}}.  The slopes with standard errors of
the best fitting regression lines in Figure~\ref{fig:scaling}, from
left to right and then top to bottom, are $-0.739$ (0.0127), 0.346
(0.0242), $-0.452$ (0.0232) and $-0.0337$ (0.0112) respectively.
These values agree with those given in
{\citeN{wei-mcw:temporal-scaling}}.

%%%%%%%%%%%%%%%%%%%%%%%%%%%%%%%%%%%%%%%%%%%%%%%%%%%%%%%%%%%%%%%%%%%%%%%%%
%%%%%%%%%%%%%%%%%%%%%%%%%%%%%%%%%%%%%%%%%%%%%%%%%%%%%%%%%%%%%%%%%%%%%%%%%

\section{Discussion}

Using a multiresolution multivariate regression model, we have been
able to accurately model and extract coherent structures from a
simulated two-dimensional fluid flow at high Reynolds number.  The
vortex statistics from analyzing each time step individually
reproduces well-known empirical scaling behavior.  This technique
allows for an efficient, objective analysis of observed turbulent
fields.  Although these results have been achieved by previous
algorithms, our method is not limited to simple accumulations of
positive or negative vorticity.  Through straightforward specification
of the template function a variety of features may be extracted from
turbulent flow.  The modular implementation of our algorithm ensures
flexibility across applications and precision of the results are
guaranteed by using sound statistical techniques.

% The next application of our multiresolution census algorithm is on
% more realistic data sets (e.g., {\shortciteNP{rup-etal:extraction}})
% -- a transfer of technology not achieved by previous methods.

The methodology presented here can be directly transferred to a
three-dimensional setting.  For freely decaying, homogeneous
geostrophic turbulence, coherent structures are compact regions of
large vorticity organized in the vertical (so-called tubes); see,
e.g., Farge {\em et al.\/} \citeyear{far-pel-sch:3D,far-pel-sch:zamm}.
Previous studies have only implemented a vortex census algorithm by
adapting the ``objective observer'' approach of
{\citeN{mcw:vortices}}.  For the multiresolution approach, software
implementations for the DWT and maximal overlap DWT are already
available that extend computations to three dimensions.  Further
collaboration with experts in turbulence would be necessary, but a
reasonable template function to try would be a three-dimensional
Gaussian kernel or possibly a cylinder of fixed radius in $(x,y)$ and
fixed height in the vertical direction.

{\shortciteN{sto-etal:asilomar}} recently proposed a novel statistical
model for object tracking using an elliptical model for coherent
structures.  A distinct feature of this tracking model is that it
allows splitting and merging of objects, and at the same time it also
allows imperfect detection of objects at each time step (e.g., false
positives).  This tracking model has been, with preliminary success,
applied to track the time evolution of coherent structures in
turbulence.  We see an opportunity to merge this technique with the
methodology presented here in order to track the time evolution of the
turbulent fluid using the regression coefficient matrix as a concise
description of the coherent structure.

% The simplification of vortex dynamics has been achieved in the past
% using point vortex models
% {\shortcite{car-etal:evolution,ben-etal:simple,wei-mcw:temporal-scaling}}.

% The ability of our method to deal with a wide variety of structures
% and in a lower signal-to-noise environment is currently under
% study. 

\section*{Acknowledgements}
The authors thank the associate editor and anonymous reviewers for
their suggestions that substantially improved the quality of this
manuscript.  This research was initiated while BW was a Visiting
Scientist in the Geophysical Statistics Project at NCAR (National
Center for Atmospheric Research).  Support for this research was
provided by the NCAR Geophysical Statistics Project, sponsored by the
National Science Foundation (NSF) under Grants DMS98-15344 and
DMS93-12686.  The work of TCML was also partially supported by the NSF
under Grant No.~0203901.

%%%%%%%%%%%%%%%%%%%%%%%%%%%%%%%%%%%%%%%%%%%%%%%%%%%%%%%%%%%%%%%%%%%%%%%%%
%\bibliography{climate,time-series,wavelet}
\bibliography{revision2nd}
\bibliographystyle{chicago}
%%%%%%%%%%%%%%%%%%%%%%%%%%%%%%%%%%%%%%%%%%%%%%%%%%%%%%%%%%%%%%%%%%%%%%%%%

\end{document}